%% file: ms.tex
\pdfoutput=1
\documentclass[11pt,a4paper,twoside,openright,closeany,textany]{book}
\usepackage{makeidx}
\usepackage{graphicx}
\usepackage{paralist}
\usepackage{caption} % to set figure caption size
\usepackage{amsmath}
\usepackage{amssymb}
\usepackage{mathtools}
\usepackage{fancyhdr}
\usepackage{multicol}
\usepackage{multirow}

\usepackage{mdframed}% http://ctan.org/pkg/mdframed
\usepackage[usenames,dvipsnames,svgnames,table]{xcolor}% http://ctan.org/pkg/xcolor
\usepackage{array}
\usepackage[%
  colorlinks=true,%
  linkcolor=black,%
  urlcolor=black,%
  citecolor=black%
]{hyperref}
\usepackage{nameref} % to enable reference to sec/chap by name
\usepackage{ifthen} % include logic
\usepackage[font={small,sl}]{caption}
\usepackage[authoryear]{natbib}
\renewcommand\bibname{References}
\newcommand{\mychapbib}{% use this at end of each chapter
  \addcontentsline{toc}{section}{\bibname}
  \bibliographystyle{natbib}
  \bibliography{strucbioinf}
}
\usepackage[colorinlistoftodos]{todonotes}
\usepackage{letltxmacro}

\usepackage{tocstyle} % to set toc raggedright
\usetocstyle{standard}
\settocfeature{raggedhook}{\raggedright}
\newcommand{\bibfont}{\small\raggedright}
\def\cite{\citep}
\newcommand{\citeeg}[1]{\cite[e.g.,][]{#1}}

\LetLtxMacro{\oldTodo}{\todo}
\renewcommand{\todo}[2][]{\oldTodo[#1]{TODO: #2}}

\usepackage[normalem]{ulem}
\newcommand\inwish[1]{\oldTodo[inline,color=SkyBlue]{WISH: #1}}

\newboolean{onechapter}

\newcommand{\AF}[1][~]{K.\@#1Anton#1Feenstra}
\newcommand{\SA}[1][~]{Sanne#1Abeln}

\newcommand{\AJ}[1][~]{Annika#1Jacobsen}

\newcommand{\BS}[1][~]{Bas#1Stringer}
\newcommand{\QH}[1][~]{Qingzhen#1Hou}

\newcommand{\OI}[1][~]{Olga#1Ivanova}

\newcommand{\JG}[1][~]{\mbox{Jose}#1\mbox{Gavald\'a-Garc\'ia}}
\newcommand{\HdF}[1][~]{\mbox{Hans}#1\mbox{de}#1\mbox{Ferrante}}

\newcommand{\KW}[1][~]{\mbox{Katharina}#1\mbox{Waury}}

\newcommand{\orcid}[1]{\href{https://orcid.org/#1}{\raisebox{-0.7ex}{\protect\includegraphics[height=3ex]{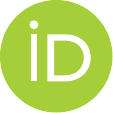}}}}
\definecolor{idgreen}{RGB}{166 206 57}
\newcommand{\mailid}[1]{\href{mailto:#1}{\raisebox{-0.3ex}{\color{idgreen}\textsf{\textbf{\Large \protect@}}}}}

\newcommand{\AFid}{\orcid{0000-0001-6755-9667}}
\newcommand{\SAid}{\orcid{0000-0002-2779-7174}}

\newcommand{\JGid}{\orcid{0000-0001-6431-3442}}

\newcommand{\OIid}{\orcid{0000-0002-9111-4593}}

\newcommand{\KWid}{\orcid{0000-0002-8570-7640}}

\newcommand{\HdFid}{\orcid{0000-0002-1772-0931}}
\newcommand{\BSid}{\orcid{0000-0001-7792-9385}}
\newcommand{\AJid}{\orcid{0000-0003-4818-2360}}

\newcommand{\QHid}{\orcid{0000-0002-9832-4518}}

\newcommand{\ACtxt}{Wrote the text}
\newcommand{\ACfig}{Created figures}
\newcommand{\ACref}{Review of current literature}
\newcommand{\ACeds}{Editorial responsibility}
\newcommand{\ACproof}{Critical proofreading}
\newcommand{\ACfb}{Non-expert feedback}

\newcommand{\Angs}[1][~]{\text{\normalfont\AA}}

\renewcommand{\and}{\quad}

\newcommand{\pdbref}[1]{\href{http://www.rcsb.org/pdb/explore.do?structureId=#1}{PDB:#1}}
\newcommand{\arxiv}[2][UNDEFINED]{\href{https://arxiv.org/abs/#2}{\ifthenelse{\equal{#1}{UNDEFINED}}{arxiv.org/abs/#2}{#1}}}

\newcommand{\figref}[2][]{\hyperref[fig:#2]{Figure\@~\ref*{fig:#2}#1}}
\newcommand{\tabref}[1]{\hyperref[tab:#1]{Table \ref*{tab:#1}}}
\renewcommand{\eqref}[2][]{\hyperref[eq:#2]{Equation#1\@~\ref*{eq:#2}}}
\newcommand{\panelref}[2][]{%
    \ifthenelse{\boolean{onechapter}}{%
        \hyperref[panel:#2]{Panel\@~``\nameref{panel:#2}#1''}%
    }{%
        \hyperref[panel:#2]{Panel\@~\ref*{panel:#2}#1}%
    }%
}
\newcommand{\secref}[2][n]{%
    \hyperref[sec:#2]{%
        \ifthenelse{\equal{#1}{n} }{Section\@~\ref*{sec:#2}}{}% just number
        \ifthenelse{\equal{#1}{nn}}{Section\@~\ref*{sec:#2} ``\nameref{sec:#2}''}{}% nm & nr
        \ifthenelse{\equal{#1}{N} }{``\nameref{sec:#2}''}{}% just quoted name
        \ifthenelse{\equal{#1}{NN} }{\nameref{sec:#2}}{}% just name
    }%
}
\newcommand{\chref}[2][n]{%
    \ifthenelse{\boolean{onechapter}}{%
        \ifthenelse{\equal{#2}{ChPref}     }{\arxiv[Chapter ``\nameref*{ch:#2}'']{1801.09442}}{}%
        \ifthenelse{\equal{#2}{ChIntroPS}  }{\arxiv[Chapter ``\nameref*{ch:#2}'']{1801.09442}}{}%
        \ifthenelse{\equal{#2}{ChDetVal}   }{\arxiv[Chapter ``\nameref*{ch:#2}'']{2108.02706}}{}%
        \ifthenelse{\equal{#2}{ChStrucAli} }{\arxiv[Chapter ``\nameref*{ch:#2}'']{1801.09442}}{}%
        \ifthenelse{\equal{#2}{ChDBClass}  }{\arxiv[Chapter ``\nameref*{ch:#2}'']{1801.09442}}{}%
        \ifthenelse{\equal{#2}{ChFunc}     }{\arxiv[Chapter ``\nameref*{ch:#2}'']{1801.09442}}{}%
        \ifthenelse{\equal{#2}{ChIntroPred}}{\arxiv[Chapter ``\nameref*{ch:#2}'']{1712.00407}}{}%
        \ifthenelse{\equal{#2}{ChHomMod}   }{\arxiv[Chapter ``\nameref*{ch:#2}'']{1712.00425}}{}%
        \ifthenelse{\equal{#2}{ChSSPred}   }{\arxiv[Chapter ``\nameref*{ch:#2}'']{1801.09442}}{}%
        \ifthenelse{\equal{#2}{ChFuncPred} }{\arxiv[Chapter ``\nameref*{ch:#2}'']{1801.09442}}{}%
        \ifthenelse{\equal{#2}{ChIntroDyn} }{\arxiv[Chapter ``\nameref*{ch:#2}'']{1801.09442}}{}%
        \ifthenelse{\equal{#2}{ChThermo}   }{\arxiv[Chapter ``\nameref*{ch:#2}'']{1801.09442}}{}%
        \ifthenelse{\equal{#2}{ChMD}       }{\arxiv[Chapter ``\nameref*{ch:#2}'']{1801.09442}}{}%
        \ifthenelse{\equal{#2}{ChMC}       }{\arxiv[Chapter ``\nameref*{ch:#2}'']{1801.09442}}{}%
    }{% else
    \hyperref[ch:#2]{%
        \ifthenelse{\equal{#1}{n} }{Chapter \ref*{ch:#2}}{}% just number
        \ifthenelse{\equal{#1}{nn}}{Chapter \ref*{ch:#2} ``\nameref{ch:#2}''}{}% name & number
        \ifthenelse{\equal{#1}{N} }{``\nameref{ch:#2}''}{}% just name
      }%
  }%
}
\newcommand{\chrefname}[1]{\hyperref[ch:#1]{Chapter \ref*{ch:#1} ``\nameref{ch:#1}''}}
\newcommand{\partref}[1]{\hyperref[#1]{Part \ref*{#1}}}
\newcommand{\appref}[1]{\hyperref[app:#1]{Appendix \ref*{app:#1}}}

\newcommand{\figsource}[1]{\protect\footnote{Figure source location: \url{#1}}}


\newenvironment{penum}[1][\itshape i)\upshape]
{\begin{inparaenum}[#1]} {\end{inparaenum}}
\newenvironment{cenum}[1][\itshape i)\upshape\ ]
{\begin{compactenum}[#1]} {\end{compactenum}}

\renewcommand{\arraystretch}{1.3}

\makeatletter \@beginparpenalty=5000 \makeatother

\newenvironment{bgreading}[1][]{
  \begin{mdframed}[%
      outerlinewidth=0,%
      linecolor=CornflowerBlue!30,%
      backgroundcolor=CornflowerBlue!30,%
      innerleftmargin=14,%
      innerrightmargin=14,%
    ]
	\ifthenelse{\equal{#1}{}}{}{% only if optional arg not empty
        \stepcounter{panel}
    	\subsection*{#1} % without panel numbers
    }
}{%
  \end{mdframed}
}

\usepackage{listings}
\definecolor{backcolour}{rgb}{0.95,0.95,0.92}
\definecolor{codegreen}{rgb}{0,0.6,0}
\definecolor{codegray}{rgb}{0.5,0.5,0.5}
\definecolor{codered}{rgb}{0.8,0,0.0}
\definecolor{codeblue}{rgb}{0.0,0,0.8}
\lstdefinestyle{codeStyle}{
    backgroundcolor=\color{backcolour},   
    commentstyle=\color{codegreen},
    keywordstyle=\color{codeblue},
    numberstyle=\tiny\color{codegray},
    stringstyle=\color{codegray},
    numbers=left,                    
    tabsize=2
} 
\makeindex

\begin{document}

\setboolean{onechapter}{true}

% page head and foots (feet):
\pagestyle{fancy}
\lhead[\small\thepage]{\small\sf\nouppercase\rightmark}
\rhead[\small\sf\nouppercase\leftmark]{\small\thepage}
\newcommand{\innerfoot}{\footnotesize{\sf{\copyright} Feenstra \& Abeln}, 2014-2023}
\newcommand{\outerfoot}{\footnotesize \sf Intro Prot Struc Bioinf}
\lfoot[\outerfoot]{\innerfoot}
\cfoot{}
\rfoot[\innerfoot]{\outerfoot}
\renewcommand{\footrulewidth}{\headrulewidth}

\mainmatter
\setcounter{chapter}{10}
\chapterauthor{\BS~\BSid \and \AJ~\AJid \and \QH~\QHid \and \HdF~\HdFid \and \OI~\OIid \and \KW~\KWid \and \JG~\JGid \and 
\SA*~\SAid~~~\AF*~\AFid}
\chapterfigure{\includegraphics[width=0.5\linewidth]{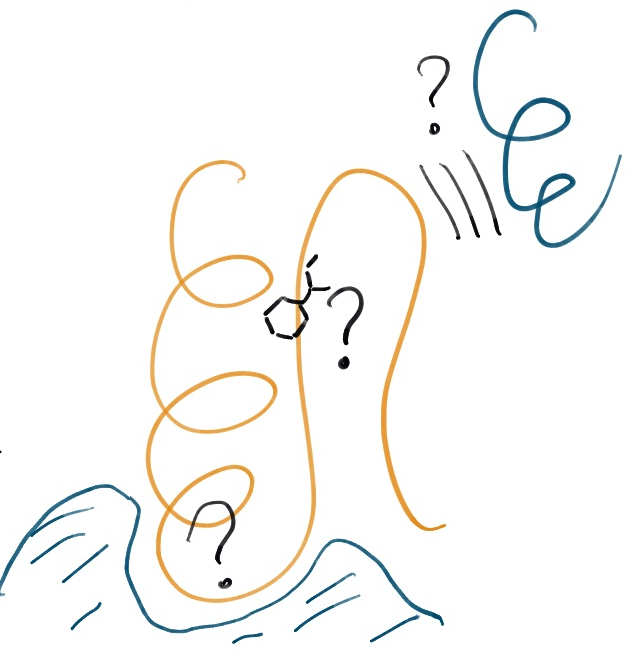}}
\chapterfootnote{* editorial responsability}
\setcounter{chapter}{10}
\chapter{Function Prediction}
\label{ch:ChFuncPred}

\ifthenelse{\boolean{onechapter}}{\tableofcontents\newpage}{}

\section{Introduction}

As mentioned in \chref{ChIntroPS}, the main motivation underlying our interest in studying protein structures is that structure relates more closely to protein function than protein sequence does. However, there are still huge gaps in our knowledge and in the mechanistic understanding of molecular function of proteins. This raises the question on how well we can predict protein function, when little to no knowledge from direct experiments is available.  

Function is a broad concept which spans different scales: from quantum scale effects for catalyzing enzymatic reactions, to phenotypes that can only be measured at the organism level, e.g.\@ comparing healthy versus diseased state. In fact, there are different research areas which focus on the elucidation of function at different scales: biochemistry at the protein level, biology at the organism level, and (bio)medicine at the level of health and disease.
In this chapter, we will consider prediction of a smaller range of functions, roughly spanning the protein residue-level up to the pathway level. We will give a conceptual overview of which functional aspects of proteins we can predict, which methods are currently available, and how well they work in practice.

\section{Different types of function prediction tasks}
Just as `function' is a really broad concept, so is the field of protein function prediction. Examples of function prediction tasks include:
\begin{compactitem}
\item cellular location for a protein;
\item which molecular pathway a protein acts in;
\item if the protein has alternative splice forms;
\item proteins regions, e.g.\@ transmembrane, and functional sites;
\item if a pair of proteins is likely to interact or bind;
\item if the protein is likely to form amyloid fibrils; % (see \chref{ChIntroPred}, 6.3.3),
\item protein stability at different temperatures, which will affect functionality;
\item of a single nucleotide polymorphism (SNP).
\end{compactitem}

\begin{figure}
\centerline{\includegraphics[width=0.8\linewidth]{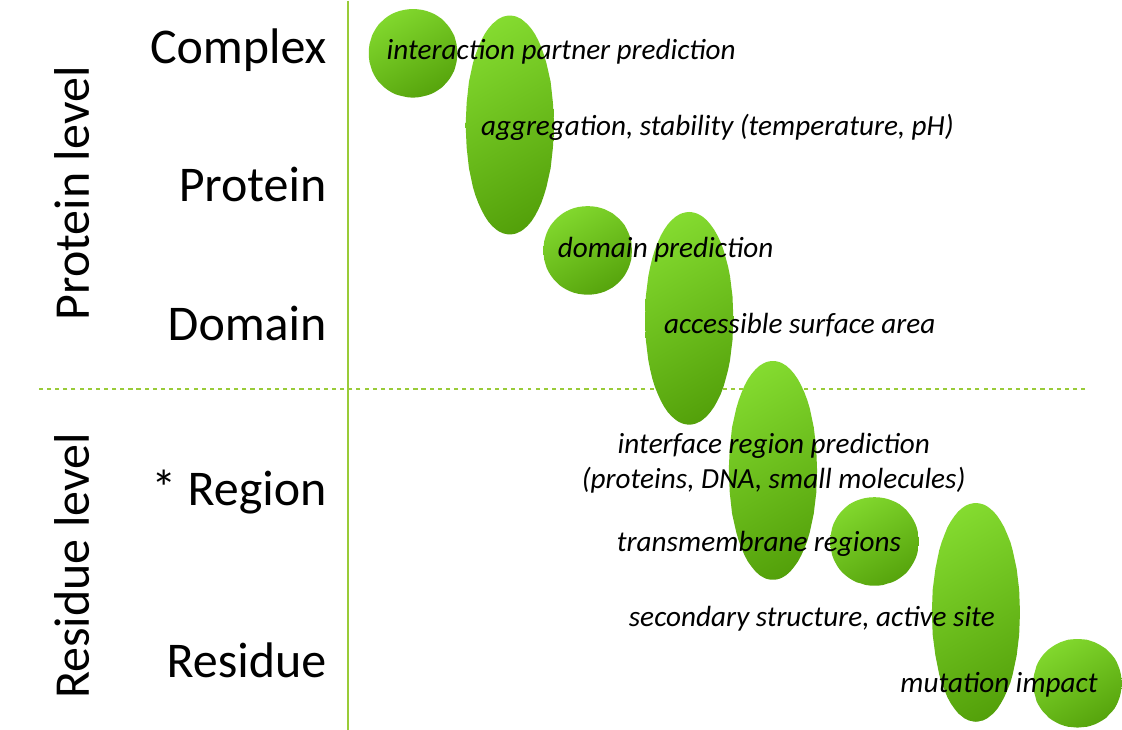}}
\textsf{* a region can be contiguous in structure, without being so in sequence.}
\caption{Protein function prediction can be performed at different levels. Level of detail goes from the top quaternary complexes, e.g.\@ proteins interacting to form a complex of multiple proteins, down to residue-level, e.g.\@ which specific amino acid residues are important for a particular function. The different types of functional features that may be predicted range from overall prediction of aggregation or stability, down to the impact of a single residue mutation.}
\label{fig:ChFuncPred-overview}
\end{figure}

We see that the purpose or output of the functional predictions acts at different levels of detail, roughly spanning residues, sequence regions, domains, proteins, complexes and pathways. Some tasks are very specific, such as prediction of the propensity to form amyloid fibrils, others are generic, such as predicting the likely functional impact upon mutation. In the next sections we will focus on three types of methods \begin{penum}
\item those that make functional predictions at residue-level and 
\item those that make functional predictions at protein-level, and 
\item those that make predictions on complexes (protein-protein interactions).
\end{penum} 
\figref{ChFuncPred-overview} gives an overview of general function prediction tasks organised according to the scale of output. 

\begin{figure}
\centerline{\includegraphics[width=0.95\linewidth]{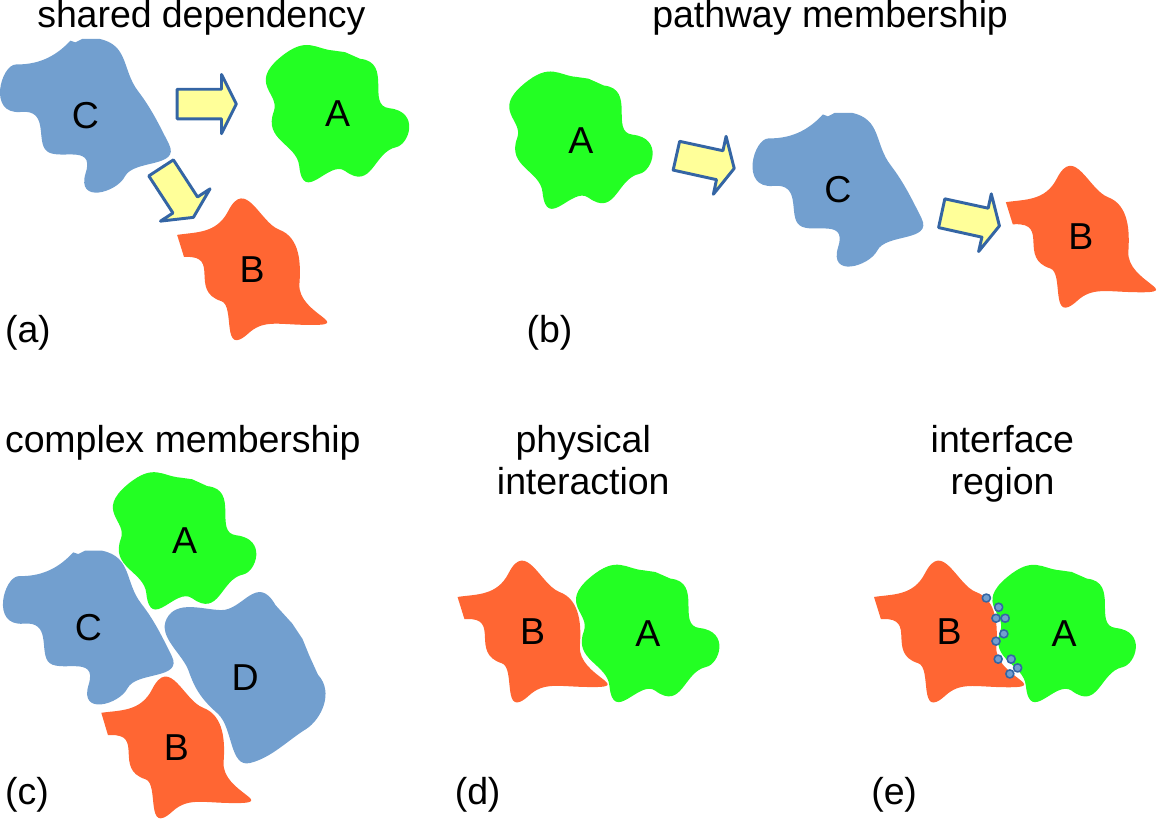}}
\caption{Overview of protein-protein interaction at different levels, and with different functional implications. (a) Mutual dependence: a correlation is observed between proteins A and B, caused by mutual dependence on protein C. (b) Indirect/cascade: the observed correlation between proteins A and B is mediated by protein C. (a) and (b) may arise through being in the same pathway. (c) Complex membership: proteins A and B are physically connected, but via intermediates C and D. (d) Direct interaction: proteins A and B are in direct physical contact. (e) The location of the interacting interface region.}
\label{fig:ChFuncPred-ppi-levels}
\end{figure}

According to \citet{Bork1998a}, protein function may be best understood in terms of protein interactions. Protein interaction may mean quite different things in different contexts, i.e.\@ at different levels. In \figref{ChFuncPred-ppi-levels} we give an overview of which types of functional relations as well as physical interactions may be captured under the notion of PPI. 

\subsection{Different function prediction methods}

Function prediction methods have some fundamental differences, both in  terms of \emph{input} and \emph{methodology}. Some methods may be able to predict function from the sequence alone, where others need homology profiles or protein structures. The underlying methodology of function prediction can span many techniques, including molecular dynamics simulations, optimisation methods, various types of machine learning, structure prediction, sequence alignment, homology searches and network analysis. 

Just as with structure prediction (see \chref[nn]{ChHomMod}), the highest accuracy for function predictions may be expected from methods that are based on homology; either by direct transfer of functional annotations from homologs, or by structure-based prediction of function based on predicted models of the structure of the proteins. However, for many proteins no additional structural or functional information is available, making it necessary to predict functional annotations based on sequence alone.

\section{Residue level function predictions}
The lowest level of protein function prediction we consider here is at the residue level. Prediction tasks that fall within this category are mutation impact analysis, active site prediction, and structural annotation predictions.
\subsection{Mutation impact analysis}

Single base-changes (mutations) in the coding regions of a protein that result in an amino acid change in the protein are known as nonsynonymous SNPs (pronounced `snips'). A single SNP can have detrimental effects for the function of a protein, but most SNPs are functionally neutral \citeeg{Mah2011}. This, and the fact that their abundance prohibits experimental analysis of all SNPs, has motivated development of a large body of bioinformatics tools that aim to predict the impact of a SNP. 

\begin{figure}
\centerline{\includegraphics[width=1.0\linewidth]{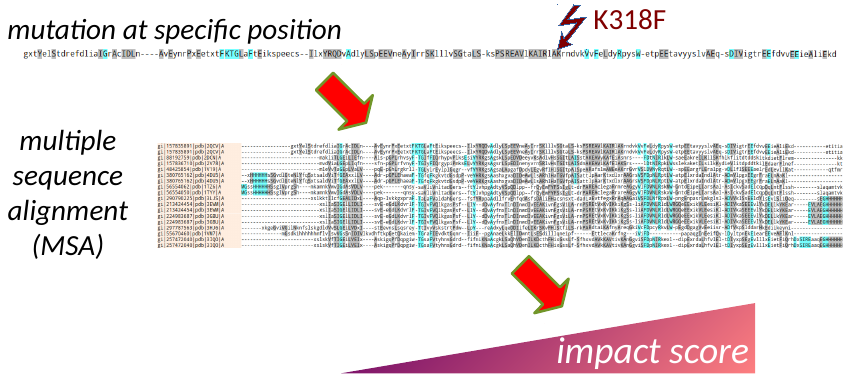}}
\caption{The concept of mutation impact prediction. From a given single amino acid change, using evolutionary information from multiple sequence alignment, one aims to assess the possible influence (impact) the mutation may have on the function of the protein.}
\label{fig:ChFuncPred-snp-impact}
\end{figure}

Most of these methods exploit that functionally relevant residues are more strongly conserved, and SNPs of these residues with physiochemically distant residues are more likely to be deleterious (see \figref{ChFuncPred-snp-impact}). Many tools also include (predicted) structural information to disentangle SNPs that affect structure from those that affect molecular function, and from thos that are unlikely to affect either structure or function. Some well-known tools are from the Baker lab \cite{Cheng2005a}, Condel \cite{Gonzalez-Perez2011}, PolyPhen-2 \cite{Adzhubei2013}, and IMHOTEP \cite{Knecht2017}; for an overview, including other tools such as SIFT, MAPP, PANTHER and MutPred, please refer to \citet{Brown2017}. Some tools also allow assessment of impact of multiple SNPs, such as PROVEAN \cite{Choi2015}, others provide a comprehensive summary with all results explained for use by e.g.\@ clinicians \cite{Venselaar2010}.

\subsection{Active site prediction}
A similarly important prediction task at the residue level is to annotate the residues in a protein that are important for the function of the protein. The goal in this task is to identify residues in the active site(s) of the protein. Here we will focus on the binding of small ligands, for example in a receptor or an enzyme. Protein-protein interactions, or PPI, and their interaction interfaces will be covered later in this chapter.

The annotation of active site binding residues in a protein can be made on the basis of three types of information \cite{Gherardini2008,Mills2015}:
\begin{penum}
\item sequence, 
\item local structure, and 
\item global structure.\end{penum} 
Methods can use only one of these sources of information, or a combination of them. Roughly speaking sequence based methods find small sequence motifs \cite{Lelieveld2016}, the local structure based methods inspects local curvature of a protein structure \cite{Zhang2011a}, and docking methods typically included an (ad-hoc) simulation of the full three-dimensional structure of the protein and ligand \cite{Lensink2016}.

\subsection{Structural annotation predictions}

Many forms of structure annotation can be used, directly or indirectly, to infer protein function. For example, the presence of a transmembrane region would give a strong suggestion on the cellular location of a protein. Similarly, disorder prediction and surface accessibility prediction may provide clues on the molecular function of a protein. Such structural feature prediction methods are covered in \chref{ChSSPred}. 

One important thing to remember is that many of these structural annotations can be predicted accurately (70-85\% accuracy), making them a reliable source of information.

\begin{bgreading}[Epitope prediction]
An epitope is a region of a protein that is recognized by the immune system; i.e.\@ it is a region to which an antibody binds. Epitope prediction is important for vaccine development, assay development for protein biomarker detection and antibody design for other purposes \cite{Sanchez-Trincado2017}. Epitope prediction may both be sequence \cite{Jespersen2017} and structure \cite{Kringelum2012, Lin2013} based. A good review of epitope prediction is \citet{Backert2015}. A recent method from our group is SeRenDIP-CE, which predicts conformational epitopes \cite{Hou2021}.

\end{bgreading}

\section{Protein level function predictions}
At the intermediate level, we can aim to predict function for a protein. Typically, such protein functions are predicted by making inferences through homology.
\subsection{Inferring function through homology}
With its three ontologies, Cellular Components, Molecular Function and Biological Process, the Gene Ontology consortium \cite{Ashburner2000,Carbon2017} aims to enable exhaustive mapping of gene function for any protein (see also \chref{ChDBClass}). Generation of such annotations however, requires costly and time-consuming experiments, and continues to be outpaced by the number of genes sequenced. To address this growing gap, researchers have aimed to automate functional annotation of proteins for already more than two decades \cite{Bork1998a}.  

A first idea for this task would be to transfer functional annotations of proteins' closest homologs identified with such annotations  \cite{Radivojac2013,Bernardes2013}. We may infer these annotations since homologous proteins are more likely to take part in the same biological process, to be located in similar cellular compartments, and to have the same or similar molecular functions. Although this approach provides a good start for identifying the protein function, it does not take into account two things: \begin{penum}
\item sequence is less conserved than structure/function, 
\item homology correlates imperfectly with functional annotations.
\end{penum} Consequently, inference of functional annotations by homology transfer alone is error prone, even with levels of sequence similarity as high as $>$60\% \cite{Rost2003}).

Often, protein-level functions are associated with a particular protein domain. When using a function prediction method it is important to realise if the methods have been developed to make predictions on a domain or full protein level. Note that there are also methods that predict the location of domains or domain boundaries; some further information is listed in the \panelref{ChIntroPred:dompred} in \chref{ChIntroPred}.

\subsection{Critical Assessment of Function Annotation}

The \textit{Critical Assessment of Functional Annotation (CAFA)} experiment is CASP's equivalent for the protein function annotation task. In this large-scale experiment, different computational methods that automate protein function annotation with Gene Ontology terms are compared (CASP is introduced in \chref{ChIntroPred}). The experiment found that the field has progressed significantly from annotation by homology transfer alone, with top-performing methods combining statistical learning with data beyond sequence similarity such as protein-protein interactions, gene expression data, and protein structure \cite{Tian2003,Radivojac2013}.

Limitations of the CAFA experiment include that quality and completeness of the GO annotations vary widely, making interpretation and usefulness of different tools dependent on the application and on which other supporting information is available \cite{Jiang2016}.

\section{Protein-protein interaction predictions}

\begin{figure}
\centerline{\includegraphics[width=1.05\linewidth]{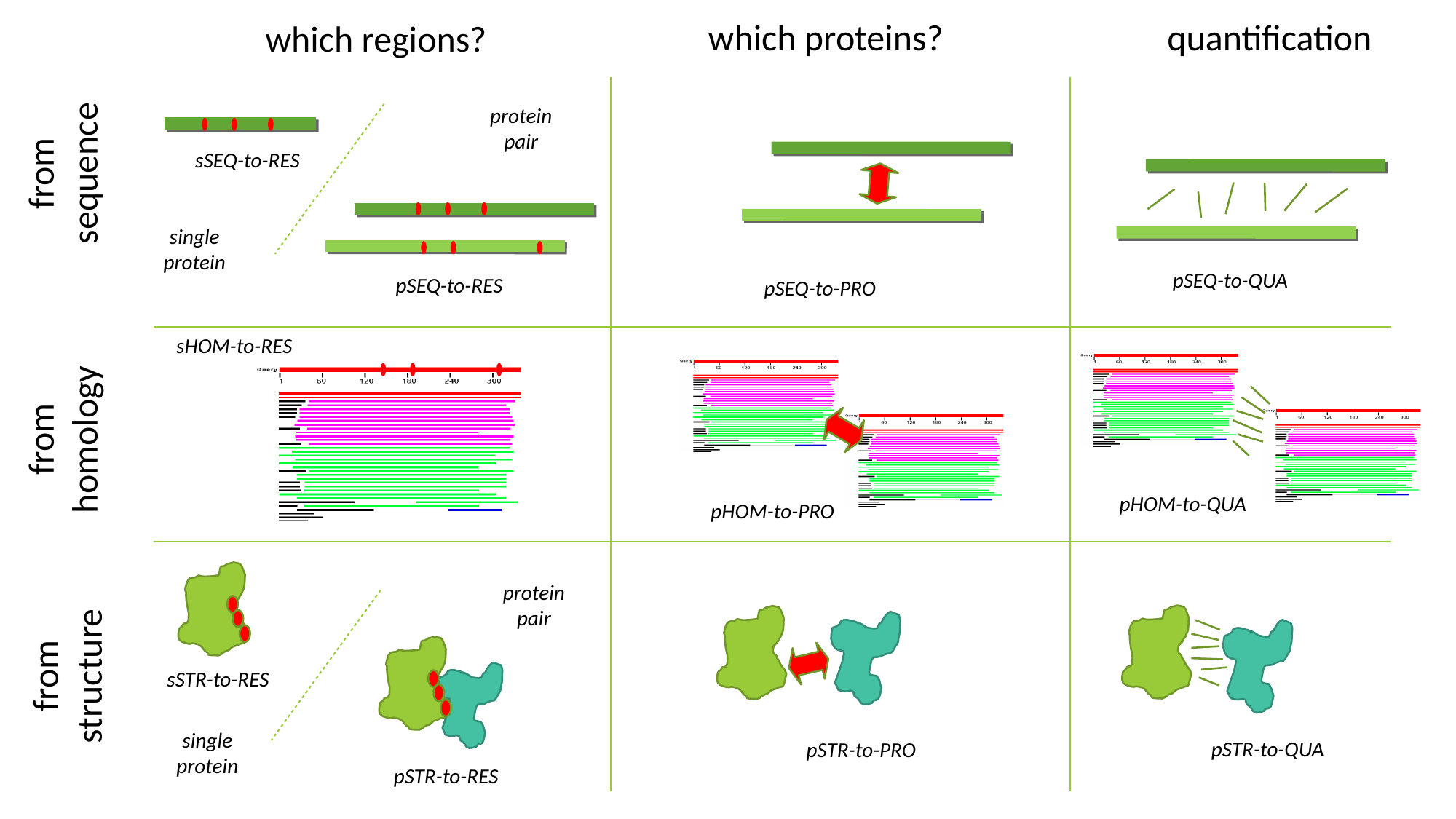}}
\caption{Levels of protein interaction prediction and types of input information. At the region level, one can predict which residues in a protein are most likely to participate in the interaction; this may be done for an individual protein without considering possible interaction partners, or for a putative interacting pair of proteins. At the protein level, one can predict which (pair of) proteins may interact, and one may furthermore quantify the interaction for example by interaction strength. Such predictions may be made from sequence data as input, from homologous sequences, or from structure data (or a combination).}
\label{fig:ChFuncPred-ppi-pred-seq-struc}
\end{figure}

Knowledge of protein-protein interactions (PPIs) can help to narrow down the biological context of proteins, e.g.\@ by suggesting in which pathways a protein is involved. PPIs may also be predicted, and such predictions can give information about function. Moreover, prediction of interaction interface, i.e.\@ the residues of either protein that make contact with the other one inside a PPI, can help to assess the impact upon mutation \cite{Ashworth2009}. An overview of this is given in \figref{ChFuncPred-ppi-pred-seq-struc}.

\subsection{Prediction of PPI from structure -- docking method}
PPI prediction, i.e.\@ the computational prediction of the complex structure of interacting proteins where the protein structures are known, is called docking. Protein-protein docking is a hard and largely unsolved problem, even though we already have the structures for both proteins. In the docking method, one strong assumption is made out of necessity: the proteins do not change their conformation; typically only side chain rearrangements in the interface region are allowed. Without this assumption, the computational cost of the predictions quickly becomes prohibitive.

Large conformational changes are hard to predict. In some way this is the same problem as for protein folding. Here, structures of homologs of the protein are used as a proxy for the possible conformations this protein could adopt, thus homology modeling is used to predict these conformations for our protein of interest. Then, regular `rigid' protein-protein docking is used. \cite{Lensink2008,Lensink2016}

\subsection{Prediction of PPI from sequence}
Prediction of PPI from sequence is an extensively studied field. Evolutionary and functional relation between proteins can be used for this purpose, since genes with closely related functions encode potentially interacting proteins. In prokaryotes functionally related proteins are often located in the same operon and thereby transcribed as a single unit. These genes can be predicted since the intergenic distance within an operon often is shorter than between operons \cite{Shoemaker2007,Juan2013}.

The phylogenetic profiling method exploits the fact that functionally related proteins, during evolution, sometimes get fused as domains into a single protein. Reversing that logic, when we observe two domains together in one protein, and we also observe homologs of the two domains as separate proteins, then we may assume these proteins are functionally related, and probably also interacting. \cite{Juan2013}.

\subsection{Protein interface prediction}
The goal of protein-interface prediction is to predict which residues constitute the interaction interface of proteins. The first task is to arrive at a definition of which residues are part of the interaction interface. Common definitions are that residues should fall below an intermolecular distance threshold, or that upon forming of the complex the accessible surface area of residues is reduced more than some threshold \cite{Esmaielbeiki2016}. 

After annotation, input of most protein-interface prediction tools is a single protein sequence (although some methods also take a pair of sequences as input) \cite{Zhang2017a}. Roughly four approaches are then taken conceptually to predict protein-protein interfaces for a given protein: 
\begin{cenum} 
\item a sequence-based where only sequence information is used to predict interface residues \cite[e.g.,][]{Murakami2010, Hou2019}, 
\item structure-based where structural information is included, 
\item a combination of sequence and structural information, and
\item template-based where known interfaces of homologous proteins are used for interface prediction \cite[e.g.,][]{Xue2011}.
\end{cenum}
The last option is by far the most reliable, if a suitable homolog complex is available. The combined option is a good second choice, if reliable structures of one or both of the interacting proteins are available \cite{Melo2016a,Esmaielbeiki2016}. 

Methods that predict PPI interface from sequence may utilize various classic Machine Learning (ML) \citep{Cheng2008, Hou2017, Hou2021} and Deep Learning (DL) architectures \citep{Shi2021, Hanson2018, Stringer2021}. Most of these methods use related structural features which are first predicted separately from sequence, such as secondary structure and solvent accessibility, as input features \cite{Ofran2007, Li2012, Hou2017}; \chref[nn]{ChSSPred} gives an overview of methods that may be used to predict these structural features. The PPI interface prediction methods use conservation \cite{Hou2017, Zhang2019}, secondary structure \cite{ Ofran2007, Zhang2019}, surface accessibility \cite{Chen2005, Hoskins2006, Zhang2019}, backbone flexibility \cite{Cilia2013,Cilia2014} or a combination of these \cite{Hou2017, Hou2019} as input features. We have recently investigated the different performances obtained from several different neural network architectures, and found that dilated convolutional networks (DCN) work well for protein interface prediction, but an ensemble network trained over the output of six other architectures (including DCN) always work best \cite{Stringer2021}. Further improvements are expected by using multi-task approaches \cite{Capel2021}.

\subsection{CAPRI}
CASP was already introduced in \chref{ChIntroPred}. In the CASP11 round, three functional aspects were explicitly scored: multimeric state, (small) ligand binding, and mutation impact. The multimeric state of proteins is which type of quarternary complex they participate in, or in other words, which and how many (other) proteins interact. These aspects were selected on being able to qualitatively evaluate them. Targets were selected that in solved crystal structure were dimeric, had a ligand bound, were from the crystallographers or in literature interest was expressed for evaluating mutants \cite{Huwe2016}. For prediction of dimer structures, only in two cases out of ten a dimer model with reasonable accuracy could be generated for the majority of monomer model structures \cite{Huwe2016}. 

A related community for the critical assessment of prediction of protein interaction (CAPRI) explicitly deals with the prediction of PPI. In the 2015 round for `easy' dimer PPI targets between 30-80\% of models generated were of `acceptable' or `medium' quality out of a top 10 models per participating predictor method. However, for harder targets (difficult dimers, multimers and heteromers), this fraction dropped to below 10\% \cite{Lensink2016}. Encouragingly, it was seen that also protein 3D structure models of lower quality could sometimes lead to acceptable or even medium quality models of the bound proteins \cite{Lensink2016}.

For ligand binding, it was found in CASP11 that the accuracies of even the best models ($\sim 2$\AA) are not good enough for accurate ligand docking \cite{Huwe2016}.
It was also the case for mutation impact prediction; for most targets, model accuracy did not correlate with accuracy of impact prediction \cite{Huwe2016}. Apparently, either homology models are not yet accurate enough for these purposes, or methods are tuned to particular characteristics of crystal structures.

\begin{bgreading}[Structure-Based Drug Design]
If we have knowledge about the three-dimensional structure of the target protein (preferably obtained through experimental methods, like crystalography) it is possible to design ligands that have a high probability of binding to it. If those ligands perform a certain task, e.g.\@ form an active complex or just the opposite - inactivate the target protein, then that ligand can be used as a drug. This is known as structure-based drug design (SBDD) \cite{Blundell1987,Blundell1996}.

While it is possible to design a new drug based only on other (known) ligand structures (ligand-based drug design), there will always be a significant level of uncertainty whether the performed comparative analysis is correct. For a broad overview of related methods, both (protein) structure-based and ligand-based, please refer to this review on Cytochrome P450 modelling by \citet{DeGraaf2005} and the more recent ones by \citet{Sliwoski2014} and by \cite{Ferreira2015}.

There are two basic approaches of designing a new drug based on a known structure. The first is a specific database search, where many potential ligands are screened, docked and scored based on how well they fit the binding site. Then, if needed, the found molecules may be modified in a desired manner and then scored again to see if they still fit. The second approach is to build a new molecule based solely on the binding site structure (its chemical and physical constraints), step by step, using a library of known fragments and applying a strategy (like growing the ligand from a ``seed'' fragment or linking best-fitting fragments). This approach has a significantly higher level of difficulty and computational complexity but allows to develop completely new molecules, not present in any database.

\end{bgreading}

\section{Key points}
\begin{compactitem}
\item Function prediction is an extremely diverse field
\item Due to large gaps in our knowledge of (molecular) functions, function prediction algorithms and techniques are in high demand.
\item Predictions may be made at residue, protein or pathway level
\item Methods can be sequence or structure based, or combined
\item Structural Bioinformatics methods such as molecular dynamics, homology recognition, sequence and structure alignment, and structural feature prediction are all used in function prediction.
\end{compactitem}

\section{Further reading}

\begin{compactitem}
\item \citet{Mah2011}: SNP impact prediction
\item \citet{Backert2015}: Epitope prediction
\item \citet{Mills2015}: Molecular function from structure
\item \citet{Ferreira2015}: Structure-based drug design
\item \citet{Jiang2016}: Critical assessment of function prediction
\end{compactitem}

\section{Author contributions}

{\renewcommand{\arraystretch}{1}
\begin{tabular}{@{}ll}
\ACtxt: &   BS, JG, AJ, QH, OI, KW \\
\ACfig: &   JG, BS, KAF \\
\ACref: &   BS, JG, KW, HdF, KAF \\
\ACproof:&  SA, AJ, HdF \\
\ACfb:  &   OI \\
\ACeds: &   SA, KAF
\end{tabular}}

\mychapbib
\clearpage
%\addcontentsline{toc}{chapter}{\bibname}
%\bibliography{strucbioinf}

\cleardoublepage

\end{document}